\documentclass[]{iptconf}

\microtypesetup{expansion=false}

\regform{
    fullname = {Глєб Шабаль},          
    birthday = {01.02.2005},                    
    position = {студент},                      
    phone = {+380 504429469},                    
    authoremail = {krivemet@ukr.net},         
    confsection = {АКТУАЛЬНІ ПИТАННЯ СУЧАСНОЇ ФІЗИКИ}, 
    copynum = {0},                              
    needliving = {ні},                          
    needinvitanion = {ні},                      
}

\usepackage[most]{tcolorbox}
\usepackage{tabularray}
\UseTblrLibrary{booktabs}
\IfFileExists{dsfont.sty}{\usepackage{dsfont}}{}
\usepackage{mathrsfs}
\usepackage{wrapfig}
\usepackage{placeins}
\usepackage{forest}
\usepackage{tikz}
\usepackage{pgfplots}
\pgfplotsset{compat=1.18}
\usepackage{listings}
\usetikzlibrary{shadows,arrows.meta,calc}

\newcommand{\be}{\begin{equation}}
\newcommand{\ee}{\end{equation}}
\newcommand{\bea}{\begin{eqnarray}}
\newcommand{\eea}{\end{eqnarray}}

\title{Equation of State of Dense Matter: 
Pauli Degeneracy, Pairing Correlations, and Implications for Neutron Stars}

\author[y.kryvenko-emetov@kpi.ua]{Y.~D.~Krivenko-Emetov}{1,2}
\author{G. Shabal}{1}

\affiliation{National Technical University of Ukraine, 03056, Kyiv}{1}
\affiliation{Institute for Nuclear Research, NAS of Ukraine, 03680, Kyiv}{2}

\udc{539.12+539.19}


\abstract{%
We develop a unified description of dense fermionic matter that consistently
incorporates Pauli degeneracy, interaction effects, and pairing correlations.
The condition that the temperature is much smaller than the Fermi energy
leads to a natural separation
between Sommerfeld, Fermi-liquid, and pairing regimes, and how these contributions
enter the equation of state. The resulting EOS is applied to the
Tolman--Oppenheimer--Volkoff equations to analyze neutron-star structure.
We demonstrate that Pauli degeneracy provides the dominant pressure,
interactions determine the stiffness of the EOS, and pairing correlations
produce subleading but potentially significant corrections, especially in
quark matter. Implications for mass--radius constraints and modern
observations are discussed. The quasiparticle interaction in the normal
Fermi-liquid phase is parameterized by the isotropic Landau parameter
$F_0^s$ and linked explicitly to the compressibility and sound speed.
The theoretical mass--radius curves are compared with four recent NICER
measurements.%
}

\begin{document}

\PaperLanguage{english}

\maketitle

\selectlanguage{english}


\pgfplotsset{compat=1.18}




\section*{Introduction}

The study of strongly interacting matter under extreme conditions remains one of the central problems of modern nuclear and high-energy physics. Experimental observations of elliptic flow in non-central heavy-ion collisions provide compelling evidence for the formation of a strongly interacting quark--gluon plasma (QGP), followed by its evolution toward a hadronic phase \cite{Fukushima2011}.

After hadronization, the system can be described as a hadron resonance gas
in the grand canonical ensemble, where conserved charges rather than
individual particle numbers determine the relevant chemical potentials. Among such approaches, the van der Waals (vdW) model provides a natural framework to incorporate short-range repulsion and intermediate-range attraction between hadrons \cite{Krivenko2017Attraction,Krivenko2019Attraction,Vovchenko2017,Krivenko2023}.
This provides an effective phenomenological description.

However, the standard vdW model is incomplete in two important aspects. First, it neglects quantum-statistical (Pauli) effects, which generate universal exchange contributions even in the absence of interactions. Second, it does not include pairing (BCS-type) correlations, which can significantly modify thermodynamic properties at low temperatures.

In this work, we develop a consistent extension of the vdW-based equation of state that incorporates both Pauli and pairing corrections and analyze their impact on the thermodynamics of hadronic and quark matter, as well as on the macroscopic structure of compact stars through the Tolman--Oppenheimer--Volkoff equations \cite{Tolman1939,Oppenheimer1939,Lattimer2001}.

In addition, the residual quasiparticle interaction in the normal Fermi
liquid is parameterized through the scalar Landau channel $F_0^s$. This
provides an explicit bridge between microscopic interaction effects,
compressibility, the sound speed, and the stellar mass--radius relation.

\section{Quantum-statistical and pairing contributions}

\subsection{Quantum-statistical (Pauli) corrections}

In addition to interaction effects, fermionic systems exhibit universal
quantum-statistical corrections due to the Pauli exclusion principle \cite{LandauLifshitz}.

In the dilute regime, the virial expansion reads with the convention
\begin{equation}
\frac{P}{T}=n+B_2(T)n^2+\mathcal O(n^3),
\end{equation}
the fermionic exchange correction is positive,
\begin{equation}
B_2^{\rm Pauli}(T)=
\frac{\lambda_T^3}{2^{5/2}g},
\end{equation}

which can be interpreted as a temperature-dependent excluded volume.

Thus, the effective virial coefficient becomes
\begin{equation}
B_{ij}^{\rm eff}(T)
=
b_{ij}
-\frac{a_{ij}}{T}
+\delta_{ij} B_i^{\rm Pauli}(T).
\end{equation}

The virial Pauli correction is applicable only in the dilute regime
$n\lambda_T^3/g\ll1$. In neutron-star interiors, where $T\ll \varepsilon_F$,
the appropriate manifestation of the Pauli principle is instead the
degenerate Fermi pressure.

\begin{table}[t]
\centering
\caption{Pauli contribution to the virial coefficient of a
spin--isospin-symmetric nucleon gas for $R_N^0=0.5$ fm and $g=4$.
Here the fourfold degeneracy combines the two spin and two isospin states.}
\small
\begin{tabular}{c c c}
\hline
$T$ (MeV) & $B_{2}^{\rm Pauli}$ (fm$^3$) & $B_{NN}^{\rm eff}$ (fm$^3$) \\
\hline
10  & 5.88  & 7.98 \\
20  & 2.08  & 4.17 \\
50  & 0.526 & 2.62 \\
100 & 0.186 & 2.28 \\
150 & 0.101 & 2.20 \\
200 & 0.066 & 2.16 \\
\hline
\end{tabular}
\label{tab:pauli_pp}
\end{table}



\begin{figure}[t]
\centering
\begin{tikzpicture}
\begin{axis}[
width=0.92\columnwidth,
height=0.62\columnwidth,
scale only axis,
xlabel={$T$ [MeV]},
ylabel={$B$ [fm$^3$]},
xmin=0, xmax=210,
ymin=0, ymax=8.5,
grid=major,
tick align=outside,
xlabel style={font=\footnotesize},
ylabel style={font=\footnotesize},
tick label style={font=\footnotesize},
legend style={
    font=\scriptsize,
    draw=none,
    fill=none,
    at={(0.97,0.97)},
    anchor=north east
},
legend cell align=left,
]

\addplot[
blue,
thick,
mark=*,
mark size=1.4pt
]
coordinates {
(10,5.88)
(20,2.08)
(50,0.526)
(100,0.186)
(150,0.101)
(200,0.066)
};
\addlegendentry{$B_{2}^{\rm Pauli}$}

\addplot[
red,
dashed,
thick
]
coordinates {
(0,2.094)
(210,2.094)
};
\addlegendentry{$b_{NN}$}

\addplot[
black,
thick,
mark=square*,
mark size=1.4pt
]
coordinates {
(10,7.98)
(20,4.17)
(50,2.62)
(100,2.28)
(150,2.20)
(200,2.16)
};
\addlegendentry{$B_{NN}^{\rm eff}$}

\end{axis}
\end{tikzpicture}
\caption{Comparison of the Pauli contribution, the classical excluded-volume
term, and the resulting effective virial coefficient of a
spin--isospin-symmetric nucleon gas.}
\label{fig:pauli_vs_vdw}
\end{figure}

\subsection{Pairing (BCS) corrections}

At temperatures $T \ll T_c$, fermionic systems undergo pairing \cite{BCS1957} ,
leading to a modification of the thermodynamic potential.

In the BCS approximation, the condensation energy is
\begin{equation}
\Delta \Omega
\simeq
-\frac{1}{2} N(0)\,\Delta^2,
\end{equation}
where $N(0)$ is the density of states at the Fermi surface for one member
of a paired time-reversed pair (equivalently, the single-spin density of
states in the standard two-component BCS problem). This convention fixes
the factor $1/2$ in the condensation energy.

At fixed chemical potential, this leads to a pressure increase
\begin{equation}
\delta P_{\rm pair}
\simeq
\frac{1}{2} N(0)\,\Delta^2,
\end{equation}
while the energy density must be obtained in a thermodynamically
consistent way from $\varepsilon=-P+\sum_i\mu_i n_i$. Its correction
therefore cannot be inferred from the sign of $\Delta\Omega$ alone.


At fixed chemical potential this increases the pressure.
Its impact on the EOS stiffness at fixed density depends on neutrality,
beta equilibrium, and the density dependence of the gap.

\subsection{Estimate of the pairing correction for quark matter}

The equation of state determines the structure of compact stars through the
Tolman--Oppenheimer--Volkoff equations.

Pairing and Pauli corrections modify the EOS:
\begin{equation}
P(\varepsilon) = P_0(\varepsilon) + \delta P(\varepsilon),
\end{equation}
which in turn affects the mass--radius relation.

To leading order, the relative change of the maximum mass scales as
\begin{equation}
\frac{\delta M_{\max}}{M_{\max}}
\sim
f_{\rm eff}
\left\langle \frac{\delta P}{P} \right\rangle,
\end{equation}
where $f_{\rm eff}$ characterizes the volume fraction of the star
affected by the correction.

\begin{table}[t]
\centering
\caption{Relative pairing correction to the degenerate quark pressure,
$\delta P_{\rm pair}/P_F = 4(\Delta/\mu_q)^2$, for representative values
of quark chemical potential $\mu_q$ and pairing gap $\Delta$.}
\label{tab:pairing_relative}
\begin{tabular}{c c c}
\hline
$\mu_q$ (MeV) & $\Delta$ (MeV) & $\delta P_{\rm pair}/P_F$ \\
\hline
400 & 25  & 0.0156 \\
400 & 50  & 0.0625 \\
400 & 75  & 0.1406 \\
400 & 100 & 0.2500 \\
\hline
450 & 25  & 0.0123 \\
450 & 50  & 0.0494 \\
450 & 75  & 0.1111 \\
450 & 100 & 0.1975 \\
\hline
500 & 25  & 0.0100 \\
500 & 50  & 0.0400 \\
500 & 75  & 0.0900 \\
500 & 100 & 0.1600 \\
\hline
550 & 25  & 0.0083 \\
550 & 50  & 0.0331 \\
550 & 75  & 0.0744 \\
550 & 100 & 0.1322 \\
\hline
600 & 25  & 0.0069 \\
600 & 50  & 0.0278 \\
600 & 75  & 0.0625 \\
600 & 100 & 0.1111 \\
\hline
\end{tabular}
\end{table}


\begin{figure}[t]
\centering
\begin{tikzpicture}
\begin{axis}[
width=0.82\columnwidth,
height=0.58\columnwidth,
scale only axis,
xlabel={$\mu_q$ [MeV]},
ylabel={$\delta P_{\rm pair}/P_F$},
xmin=390, xmax=610,
ymin=0, ymax=0.27,
grid=major,
tick align=outside,
xlabel style={font=\scriptsize},
ylabel style={font=\scriptsize},
tick label style={font=\scriptsize},
legend style={
    font=\tiny,
    draw=none,
    fill=white,
    fill opacity=0.75,
    text opacity=1,
    at={(0.97,0.97)},
    anchor=north east
},
legend cell align=left,
]

\addplot[thick, mark=*, mark size=1.1pt] coordinates {
(400,0.0156) (450,0.0123) (500,0.0100) (550,0.0083) (600,0.0069)
};
\addlegendentry{$\Delta=25$ MeV}

\addplot[thick, mark=square*, mark size=1.1pt] coordinates {
(400,0.0625) (450,0.0494) (500,0.0400) (550,0.0331) (600,0.0278)
};
\addlegendentry{$\Delta=50$ MeV}

\addplot[thick, mark=triangle*, mark size=1.2pt] coordinates {
(400,0.1406) (450,0.1111) (500,0.0900) (550,0.0744) (600,0.0625)
};
\addlegendentry{$\Delta=75$ MeV}

\addplot[thick, mark=diamond*, mark size=1.2pt] coordinates {
(400,0.2500) (450,0.1975) (500,0.1600) (550,0.1322) (600,0.1111)
};
\addlegendentry{$\Delta=100$ MeV}

\end{axis}
\end{tikzpicture}
\caption{Relative pairing correction to the degenerate quark pressure as a function of quark chemical potential for several values of the pairing gap.}
\label{fig:pairing_relative}
\end{figure}

For ultrarelativistic quark matter, the relative pairing correction to the
degenerate Fermi pressure is
\begin{equation}
\frac{\delta P_{\rm pair}}{P_F}=4\left(\frac{\Delta}{\mu_q}\right)^2.
\end{equation}

For these estimates we use the standard CFL result
\cite{Alford2008,Kurkela2024}.
A more detailed pedagogical treatment of the combined Pauli-degeneracy,
vdW-interaction, and BCS-pairing framework, including these numerical
estimates for quark matter, is also given in the bachelor's thesis of
 \cite{ShabalBachelor2026}.

Thus, the effect decreases as $\mu_q^{-2}$ and grows with $\Delta^2$.
For $\mu_q=400$--$600$ MeV and $\Delta=25$--$100$ MeV, one finds
\begin{equation}
\frac{\delta P_{\rm pair}}{P_F}\sim 0.7\%\text{--}25\%,
\end{equation}
which shows that color-superconducting pairing can provide a substantial
correction to the quark-matter equation of state.

\subsection{Simultaneous inclusion of Pauli degeneracy and pairing correlations in quark matter}

For cold quark matter in neutron-star interiors, the Pauli principle and
pairing correlations must be incorporated simultaneously at the level of the
degenerate equation of state.

In the ultrarelativistic limit, the Pauli principle generates \cite{Fukushima2011}  the dominant
degeneracy pressure,
\begin{equation}
P_F=\frac{g\mu_q^4}{24\pi^2(\hbar c)^3},
\qquad
\varepsilon_F=3P_F,
\qquad
n_B=\frac{g\mu_q^3}{18\pi^2(\hbar c)^3},
\end{equation}
where $g=18$ for three colors and three light flavors.

Pairing correlations in the color-superconducting phase
\cite{Alford2008,Kurkela2024} produce an additional
condensation contribution
\begin{equation}
\delta P_{\rm pair}\simeq
3\,\frac{\mu_q^2\Delta^2}{\pi^2(\hbar c)^3},
\qquad
\delta\varepsilon_{\rm pair}
=-\delta P_{\rm pair}
+\mu_q\frac{\partial\delta P_{\rm pair}}{\partial\mu_q}.
\end{equation}

For a density-independent gap over the interval considered, this gives
$\delta\varepsilon_{\rm pair}=+\delta P_{\rm pair}$. If
$\Delta=\Delta(\mu_q)$, the derivative must also act on the gap; hence the
energy-density correction cannot in general be assigned from the sign of
the condensation energy alone.

Thus, the total equation of state can be written as
\begin{equation}
P_{\rm tot}(\mu_q)=
\frac{g\mu_q^4}{24\pi^2(\hbar c)^3}
+
3\,\frac{\mu_q^2\Delta^2}{\pi^2(\hbar c)^3}
-
B_{\rm bag},
\end{equation}
\begin{equation}
\varepsilon_{\rm tot}(\mu_q)=
\frac{g\mu_q^4}{8\pi^2(\hbar c)^3}
+
3\,\frac{\mu_q^2\Delta^2}{\pi^2(\hbar c)^3}
+
B_{\rm bag}.
\end{equation}

The last expression assumes that $\Delta$ is constant over the displayed
chemical-potential interval. For a density-dependent gap, it acquires the
additional term
$6\mu_q^3\Delta\,d\Delta/d\mu_q\,[\pi^2(\hbar c)^3]^{-1}$.

The relative importance of pairing with respect to the Pauli degeneracy pressure is
\begin{equation}
\frac{\delta P_{\rm pair}}{P_F}
=
\frac{72}{g}\frac{\Delta^2}{\mu_q^2}
=
4\frac{\Delta^2}{\mu_q^2},
\qquad (g=18).
\end{equation}

Therefore, in quark cores of neutron stars the Pauli principle provides the
dominant background stiffness of the EOS, while pairing correlations produce
a substantial correction that can reach the level of several percent and,
for large gaps, up to $\mathcal{O}(10\%)$--$\mathcal{O}(20\%)$.

\begin{table}[t]
\centering
\caption{Representative estimates for quark matter with simultaneous Pauli and pairing contributions.}
\label{tab:pauli_pairing_compact}
\begin{tabular}{c c c c c}
\hline
$\mu_q$ & $\Delta$ & $n_B$ & $P_F$ & $\delta P_{\rm pair}/P_F$ \\
(MeV) & (MeV) & (fm$^{-3}$) & (MeV fm$^{-3}$) &  \\
\hline
400 & 50  & 0.844 & 253  & 0.063 \\
450 & 50  & 1.20  & 406  & 0.049 \\
450 & 100 & 1.20  & 406  & 0.197 \\
500 & 100 & 1.65  & 618  & 0.160 \\
550 & 100 & 2.19  & 905  & 0.132 \\
600 & 100 & 2.85  & 1281 & 0.111 \\
\hline
\end{tabular}
\end{table}

Table~\ref{tab:pauli_pairing_compact} shows representative numerical estimates
for cold quark matter when the Pauli degeneracy pressure and pairing correlations
are included simultaneously. One sees that the Pauli principle provides the dominant
background stiffness of the equation of state through the degenerate Fermi pressure
$P_F\propto \mu_q^4$, whereas the pairing contribution produces an additional
positive correction of order $\delta P_{\rm pair}\propto \mu_q^2\Delta^2$.

For moderate gaps, $\Delta\sim 50$ MeV, the pairing contribution remains at the
level of several percent of $P_F$. For larger gaps, $\Delta\sim 100$ MeV, the effect
can reach the level of $\sim 10\%$--$20\%$, which is large enough to noticeably
modify the equation of state and, consequently, the TOV mass--radius relation.

\section{Low-temperature regimes of dense fermionic matter in neutron stars}
\label{sec:lowT_hierarchy}

The thermodynamic properties of dense matter in neutron-star interiors
are controlled by the hierarchy between the temperature $T$, the Fermi
energy $\varepsilon_F$, and, when present, the pairing gap $\Delta$ and
critical temperature $T_c$. In realistic neutron stars one typically has
\begin{equation}
T \ll \varepsilon_F,
\end{equation}
which places the system deep in the degenerate regime. In this section,
we summarize the appropriate theoretical descriptions in different
temperature domains.

\subsection{Degenerate regime: \texorpdfstring{$T \ll \varepsilon_F$}{T << epsilon_F}}

For all relevant components (electrons, nucleons, and possibly quarks),
the Fermi energy is of order
\begin{equation}
\varepsilon_F \sim 10\text{--}100~\mathrm{MeV}
\end{equation}
for nucleons and up to several hundred MeV for quarks, while the stellar
temperature satisfies
\begin{equation}
T \sim 10^{-2}\text{--}10^{-1}~\mathrm{MeV}.
\end{equation}
Thus,
\begin{equation}
\frac{T}{\varepsilon_F} \ll 1.
\end{equation}

In this regime, thermodynamic quantities admit the Sommerfeld expansion. For a generic observable
\begin{equation}
I(T)=\int_0^\infty d\epsilon\, \phi(\epsilon)\,
\frac{1}{e^{(\epsilon-\mu)/T}+1},
\end{equation}
one obtains
\begin{equation}
I(T)=\int_0^{\mu}\phi(\epsilon)\,d\epsilon
+
\frac{\pi^2}{6}T^2\phi'(\mu)
+\mathcal O(T^4).
\end{equation}

As a consequence, the equation of state (EOS) takes the form
\begin{equation}
P(T,\mu)=P_0(\mu)+\mathcal O(T^2),
\qquad
\varepsilon(T,\mu)=\varepsilon_0(\mu)+\mathcal O(T^2),
\end{equation}
so that the zero-temperature EOS provides the dominant contribution.

In this regime, the Pauli principle manifests itself through the degeneracy pressure

\begin{equation}
P_F \propto \mu^4,
\end{equation}
rather than through the virial expansion valid for dilute gases.

\subsection{Interacting normal phase: Landau Fermi liquid}

For nucleonic matter, the interparticle interaction is strong.
Nevertheless, at low temperatures the system can be described within
the framework of Landau Fermi-liquid theory.

In this approach, the system is represented as a gas of quasiparticles
with dispersion relation $\varepsilon(p)$ near the Fermi surface.
The effects of interactions are encoded in:

\begin{itemize}
\item the effective mass $m^*$,
\item the Landau parameters $F_\ell$, $G_\ell$, \dots,
\item the renormalized density of states $N^*(0)$.
\end{itemize}

The quasiparticle interaction may be introduced directly through the
Landau energy functional,
\begin{equation}
\delta E
=
\sum_{\mathbf p\sigma}
\varepsilon_{\mathbf p\sigma}\,\delta n_{\mathbf p\sigma}
+
\frac{1}{2}
\sum_{\mathbf p\sigma,\mathbf p'\sigma'}
f_{\sigma\sigma'}(\mathbf p,\mathbf p')
\delta n_{\mathbf p\sigma}\delta n_{\mathbf p'\sigma'}.
\label{eq:landau_energy_functional}
\end{equation}
Near the Fermi surface, the dimensionless interaction is expanded in
Legendre polynomials:
\begin{equation}
N^*(0)f_{\sigma\sigma'}(\cos\theta)
=
\sum_{\ell=0}^{\infty}
\left[
F_\ell^s+F_\ell^a\,
\boldsymbol{\sigma}\!\cdot\!\boldsymbol{\sigma}'
\right]
P_\ell(\cos\theta).
\label{eq:landau_expansion}
\end{equation}
For an isotropic variation of the baryon density, the scalar channel
$F_0^s$ is the relevant one. If
$N^*(0)=\partial n_B/\partial\varepsilon_F^*$ contains the appropriate
degeneracy factor, then
\begin{equation}
\frac{\partial\mu_B}{\partial n_B}
=
\frac{1+F_0^s}{N^*(0)},
\qquad
\frac{dP_{\rm FL}}{dn_B}
=
\frac{n_B[1+F_0^s(n_B)]}{N^*(0;n_B)}.
\label{eq:landau_compressibility}
\end{equation}
The interaction contribution is therefore not merely a formal additive
term:
\begin{equation}
\begin{aligned}
P_{\rm FL}(n_B)
&=
\int_0^{n_B}dn\,
\frac{n[1+F_0^s(n)]}{N^*(0;n)},\\
P_{\rm int}^{(L)}&=P_{\rm FL}-P_F.
\end{aligned}
\label{eq:landau_pressure}
\end{equation}
When $m^*$ and $F_0^s$ vary slowly, this gives
$P_{\rm FL}\simeq(1+F_0^s)P_F$. Mechanical stability requires
$F_0^s>-1$, while in natural units the sound speed is
\begin{equation}
c_s^2
=
\frac{dP}{d\varepsilon}
=
\frac{n_B}{\mu_B}
\frac{1+F_0^s}{N^*(0)},
\qquad 0<c_s^2\leq1.
\label{eq:landau_sound_speed}
\end{equation}
For multicomponent $npe\mu$ matter, $F_0^s$ is generalized to a matrix
of parameters $F_0^{ij}$. Below we use a one-component effective
parameterization, sufficient for estimating the change in stiffness.
Fermi-liquid theory, Landau parameters, and their relation to the EOS and
the pairing gap have already been studied for neutron matter
\cite{Schwenk2003,Alp2025}. The contribution of the present work is
therefore not the first application of this formalism to stars, but its
explicit incorporation into the unified ``Pauli degeneracy--interaction--
pairing'' framework and the comparison of that framework with observations.

The thermodynamic quantities retain the same functional structure as in
the ideal Fermi gas, but with renormalized parameters. In particular,
the pressure may be schematically written as
\begin{equation}
P = P_F(m^*) + P_{\rm int},
\end{equation}
where $P_F$ represents the degeneracy (Pauli) contribution and
$P_{\rm int}$ encodes interaction effects.

Thus, the statement that dense matter becomes ``more ideal'' at high
density should be understood in the sense that the low-energy excitations
are well described by weakly interacting quasiparticles, rather than
that the underlying interaction is weak.

\subsection{Superfluid phase: \texorpdfstring{$T<T_c$}{T < T_c}}

At sufficiently low temperatures, attractive channels in the interaction
lead to Cooper pairing and the formation of a superfluid or superconducting
phase \cite{Schwenk2003,SedrakianClark2019}. In the weak-coupling BCS
approximation, the critical temperature is related to the pairing gap as
\cite{BCS1957}

\begin{equation}
T_c \simeq 0.57\,\Delta(0).
\end{equation}

For neutron-star matter, microscopic calculations predict primarily
neutron ${}^{1}S_0$ pairing in the inner crust, proton ${}^{1}S_0$
pairing, and neutron ${}^{3}P_2$--${}^{3}F_2$ pairing in the core
\cite{SedrakianClark2019}. Cooling observations of the neutron star in
Cassiopeia~A have been used to constrain the density-dependent critical
temperature of triplet neutron pairing. Within the standard
pair-breaking-and-formation interpretation, a characteristic maximum
$T_{Cn}^{\max}\sim(5$--$10)\times10^8$~K is obtained
\cite{Shternin2021}.

In the paired phase, the thermodynamic potential acquires an additional
condensation contribution,
\begin{equation}
\Delta \Omega \simeq -\frac{1}{2}N(0)\Delta^2,
\end{equation}
which leads to a correction to the pressure at fixed chemical potential,
\begin{equation}
\delta P_{\rm pair} \simeq \frac{1}{2}N(0)\Delta^2.
\end{equation}

The total pressure can therefore be written as
\begin{equation}
P = P_F + P_{\rm int} + \delta P_{\rm pair}.
\end{equation}

At fixed chemical potential the pairing contribution increases the pressure.
At fixed density, however, its effect on the EOS stiffness depends on the
self-consistent adjustment of chemical potentials under neutrality and
beta-equilibrium conditions.

In contrast to the normal phase, low-temperature corrections are no longer
polynomial in $T$, but instead become exponentially suppressed,
\begin{equation}
\delta P(T) \sim e^{-\Delta/T}.
\end{equation}

\subsection{Hierarchy of regimes}

The above discussion leads to the following hierarchy of physical regimes:

\begin{itemize}
\item \textbf{Dilute regime:} $n\lambda_T^3/g \ll 1$ \\
Virial expansion applies; Pauli effects appear as corrections to $b_2$.

\item \textbf{Degenerate regime:} $T \ll \varepsilon_F$ \\
Sommerfeld expansion; Pauli principle manifests as degeneracy pressure.

\item \textbf{Interacting regime:} $T \ll \varepsilon_F$ \\
Landau Fermi-liquid description; interactions encoded in quasiparticles.

\item \textbf{Paired regime:} $T < T_c$ \\
Superfluid/superconducting phase; pairing corrections $\propto \Delta^2$.
\end{itemize}

\subsection{Physical interpretation}

In neutron-star matter, the pressure is dominated by Pauli degeneracy.
Strong interactions modify this background via renormalization of
quasiparticle properties, while pairing correlations provide an additional,
typically subleading but non-negligible correction.

Thus, the equation of state of dense matter can be viewed as a superposition
of three physically distinct contributions:
\begin{equation}
\boxed{
P = P_{\rm Pauli} + P_{\rm interaction} + P_{\rm pairing}.
}
\end{equation}


\subsection{Connection between the low-temperature hierarchy, the TOV
equations, and the mass--radius relation}

The hierarchy of temperature scales discussed above determines how
microscopic physics enters the stellar-structure problem. In the
Tolman--Oppenheimer--Volkoff equations, all contributions act through a
single thermodynamically consistent equation of state,
\begin{equation}
P(\varepsilon)=P_{\mathrm{Pauli}}(\varepsilon)
+P_{\mathrm{int}}(\varepsilon)
+\delta P_{\mathrm{pair}}(\varepsilon).
\label{eq:eos_hierarchy_bridge}
\end{equation}
The TOV system itself and its boundary conditions are given once below in
Sec.~\ref{sec:tov_fl}; here it is sufficient to distinguish thermal,
interaction-induced, and pairing-induced changes of the EOS.

\paragraph{Degenerate regime: \texorpdfstring{$T\ll\varepsilon_F$}{T << epsilon_F}.}

For mature neutron stars, $T/\varepsilon_F\ll1$, so the correction
$\delta P_T=\mathcal O(T^2)$ is small and the approximation
$P(\varepsilon,T)\simeq P_0(\varepsilon)$ is usually sufficient for
the global structure.

\paragraph{Interacting normal phase: Landau Fermi liquid.}

In the normal phase, interactions renormalize the quasiparticle spectrum
and the compressibility. For stellar structure, the decisive quantity is
the total stiffness $dP/d\varepsilon$: increasing it strengthens the
resistance to gravitational compression and generally increases both the
radius and the maximum mass.

\paragraph{Paired phase: \texorpdfstring{$T<T_c$}{T < T_c}.}

At $T<T_c$, the gap suppresses fermionic excitations and the condensation
term modifies the effective zero-temperature EOS. In the BCS estimate its
scale is $\delta P_{\mathrm{pair}}\simeq N(0)\Delta^2/2$, as given above.

\paragraph{Impact on TOV solutions.}

The hierarchy leads to a clear ordering of structural effects:
\begin{enumerate}
\item The Pauli principle generates the dominant degeneracy pressure and
sets the basic scale of stellar support against gravity.

\item Strong interactions modify the compressibility and symmetry
properties of dense matter, thereby determining the detailed stiffness of
the EOS and the shape of the mass--radius curve.

\item Pairing correlations provide an additional pressure correction that
is usually small for nucleonic superfluidity but can be appreciable in
quark matter.
\end{enumerate}

Thus, the mass--radius relation is governed by the effective sequence
\begin{equation}
P_{\mathrm{Pauli}}
\Longrightarrow
P_{\mathrm{Pauli}}+P_{\mathrm{int}}
\Longrightarrow
P_{\mathrm{Pauli}}+P_{\mathrm{int}}+\delta P_{\mathrm{pair}}.
\end{equation}

\paragraph{Scaling estimate for \texorpdfstring{$M(R)$}{M(R)}.}

If the EOS corrections are small,
\begin{equation}
|\delta P|\ll P_0,
\qquad
|\delta\varepsilon|\ll\varepsilon_0,
\end{equation}
their impact on TOV solutions may be estimated in the simplest
perturbative approximation:
\begin{equation}
\frac{\delta R}{R}
\sim
\left\langle\frac{\delta P}{P_0}\right\rangle_{\mathrm{core}},
\qquad
\frac{\delta M}{M}
\sim
f_{\mathrm{eff}}
\left\langle\frac{\delta P}{P_0}\right\rangle_{\mathrm{core}}.
\end{equation}
It follows directly that
\begin{itemize}
\item Sommerfeld corrections of order $T^2$ are practically negligible;
\item Fermi-liquid renormalizations significantly affect the entire
mass--radius curve;
\item in the weak-coupling nucleonic case, pairing usually produces only a
small correction to global $M$ and $R$, although it strongly affects
cooling and transport \cite{SedrakianClark2019,Shternin2021};
\item color superconductivity can noticeably change the stellar radius
and maximum mass \cite{Alford2008,Kurkela2024}.
\end{itemize}

\section{Equation of state, stellar structure, and observational constraints}

\subsection{Baseline mass--radius curve and observational benchmarks}
\label{sec:mr_uleiev}

\begin{figure}[ht]
\centering
\begin{tikzpicture}
\node[anchor=south west,inner sep=0] (mrimage) at (0,0)
{\includegraphics[width=\columnwidth]{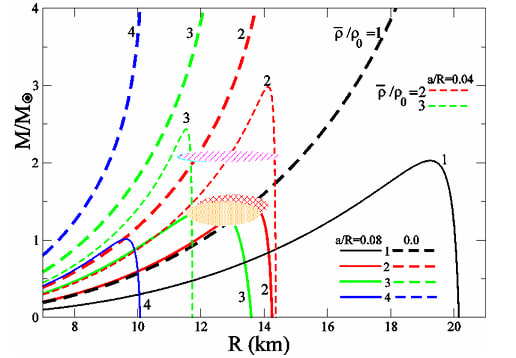}};
\begin{scope}[x={(mrimage.south east)},y={(mrimage.north west)}]
\draw[magenta!75!black,line width=0.65pt]
  (0.365163,0.388144)--(0.509040,0.388144)
  (0.428836,0.349294)--(0.428836,0.433850);
\draw[magenta!75!black,line width=0.65pt]
  (0.365163,0.3815)--(0.365163,0.3945)
  (0.509040,0.3815)--(0.509040,0.3945)
  (0.4223,0.349294)--(0.4353,0.349294)
  (0.4223,0.433850)--(0.4353,0.433850);
\filldraw[fill=magenta!75!black,draw=white,line width=0.35pt]
  (0.428836,0.388144) circle (1.55pt);
\node[font=\tiny,fill=white,fill opacity=0.82,text opacity=1,
      inner sep=0.6pt,anchor=south west]
  at (0.440,0.411) {J0030};
\draw[cyan!65!black,line width=0.65pt]
  (0.363326,0.535087)--(0.495571,0.535087)
  (0.417204,0.519319)--(0.417204,0.550856);
\draw[cyan!65!black,line width=0.65pt]
  (0.363326,0.5285)--(0.363326,0.5415)
  (0.495571,0.5285)--(0.495571,0.5415)
  (0.4107,0.519319)--(0.4237,0.519319)
  (0.4107,0.550856)--(0.4237,0.550856);
\filldraw[fill=cyan!65!black,draw=white,line width=0.35pt]
  ($(0.417204,0.535087)+(-1.5pt,-1.5pt)$)
  rectangle
  ($(0.417204,0.535087)+(1.5pt,1.5pt)$);
\node[font=\tiny,fill=white,fill opacity=0.82,text opacity=1,
      inner sep=0.6pt,anchor=south west]
  at (0.427,0.556) {J0740};
\draw[orange!90!black,line width=0.65pt]
  (0.309448,0.385402)--(0.406183,0.385402)
  (0.348020,0.376946)--(0.348020,0.393857);
\draw[orange!90!black,line width=0.65pt]
  (0.309448,0.3789)--(0.309448,0.3919)
  (0.406183,0.3789)--(0.406183,0.3919)
  (0.3415,0.376946)--(0.3545,0.376946)
  (0.3415,0.393857)--(0.3545,0.393857);
\filldraw[fill=orange!90!black,draw=white,line width=0.35pt]
  ($(0.348020,0.385402)+(0,1.9pt)$)--
  ($(0.348020,0.385402)+(-1.7pt,-1.2pt)$)--
  ($(0.348020,0.385402)+(1.7pt,-1.2pt)$)--cycle;
\node[font=\tiny,fill=white,fill opacity=0.82,text opacity=1,
      inner sep=0.6pt,anchor=north east]
  at (0.337,0.361) {J0437};
\draw[green!55!black,line width=0.65pt]
  (0.229857,0.390429)--(0.344346,0.390429)
  (0.282510,0.374433)--(0.282510,0.404141);
\draw[green!55!black,line width=0.65pt]
  (0.229857,0.3839)--(0.229857,0.3969)
  (0.344346,0.3839)--(0.344346,0.3969)
  (0.2760,0.374433)--(0.2890,0.374433)
  (0.2760,0.404141)--(0.2890,0.404141);
\filldraw[fill=green!55!black,draw=white,line width=0.35pt]
  ($(0.282510,0.390429)+(0,2.0pt)$)--
  ($(0.282510,0.390429)+(-2.0pt,0)$)--
  ($(0.282510,0.390429)+(0,-2.0pt)$)--
  ($(0.282510,0.390429)+(2.0pt,0)$)--cycle;
\node[font=\tiny,fill=white,fill opacity=0.82,text opacity=1,
      inner sep=0.6pt,anchor=north east]
  at (0.270,0.419) {J0614};
\end{scope}
\end{tikzpicture}
\caption{The baseline neutron-star mass--radius relations are reproduced
from Fig.~3 of Uleiev et al.~\cite{Uleiev2026}. Solid curves contain both
volume and surface terms, whereas dashed curves represent the volume
approximation; curve families are specified by $a/R$ and
$\bar\rho/\rho_0$. The shaded regions also belong to the source figure and
show the observational constraints used by its authors. Colored markers
with asymmetric error bars are added in the present work and show
one-dimensional projections of the $68\%$ NICER regions for
PSR~J0030+0451, PSR~J0740+6620, PSR~J0437-4715, and PSR~J0614-3329 from
the primary analyses
\cite{Kini2026,Salmi2024,Choudhury2024,Mauviard2025}. The cross-shaped
intervals do not reproduce the correlation between $M$ and $R$ within the
full two-dimensional posterior. An explicit Landau correction and a
superconducting-pairing contribution were not part of the stated level of
the baseline construction and are considered here as natural extensions.}
\label{fig:CSC_MR}
\end{figure}

The baseline mass--radius curves in Fig.~\ref{fig:CSC_MR} are reproduced
from Fig.~3 of Uleiev et al.~\cite{Uleiev2026}. That work provides a
successful, physically transparent, and analytically controlled
macroscopic description of a cold, slowly rotating neutron star in the
effective-surface approximation. An important strength is the consistent
combination of volume and surface contributions in a leptodermic expansion,
together with rotation in a unified scheme. The density functional is
specified by the incompressibility $K$, effective mass $m$, and gradient
coefficient $C$; the latter controls, among other quantities, the
characteristic crust thickness $a$. With only a small number of physically
interpretable parameters, the model clearly demonstrates how surface
properties and interior density shape families of $M(R)$ curves.

Interactions are not neglected in the approach of Ref.~\cite{Uleiev2026}:
their mean-field and surface effects are incorporated phenomenologically in
$K$ and $C$. The authors did not introduce a microscopic Landau
quasiparticle interaction function, a separate $F_0^s$ parameter, or a
pairing gap $\Delta$, because those effects were outside the stated
macroscopic level of description. This separation of scales is useful: it
allows their model to serve as a well-defined baseline to which residual
Fermi-liquid and pairing corrections can be added consistently without
double counting. The correlation between the coordinates $t$ and $\varphi$
discussed in their moment-of-inertia calculation describes a geometric
rotation effect and is physically distinct from Cooper pairing.

The curves of Uleiev et al. therefore provide a valuable macroscopic
benchmark for subsequent microscopic refinements. Their solid and dashed
branches systematically display the roles of the relative crust thickness
$a/R$ and interior density $\bar\rho/\rho_0$. In the original work, the
comparison used the PSR~J0030+0451 and PSR~J0740+6620 data available at the
time of preparation.

For an expanded comparison, we use the $68\%$ intervals for
PSR~J0030+0451~\cite{Kini2026}, PSR~J0740+6620~\cite{Salmi2024},
PSR~J0437-4715~\cite{Choudhury2024}, and the recent result for
PSR~J0614-3329~\cite{Mauviard2025}. The first three are also summarized in
Ref.~\cite{Dexheimer2026}:
\begin{equation}
\begin{aligned}
\mathrm{PSR~J0030{+}0451}:\;&
M=1.43^{+0.20}_{-0.17}M_\odot,\\[-1mm]
&R=12.68^{+1.31}_{-1.04}\ \mathrm{km},\\
\mathrm{PSR~J0740{+}6620}:\;&
M=2.073\pm0.069M_\odot,\\[-1mm]
&R=12.49^{+1.28}_{-0.88}\ \mathrm{km},\\
\mathrm{PSR~J0437{-}4715}:\;&
M=1.418\pm0.037M_\odot,\\[-1mm]
&R=11.36^{+0.95}_{-0.63}\ \mathrm{km},\\
\mathrm{PSR~J0614{-}3329}:\;&
M=1.44^{+0.06}_{-0.07}M_\odot,\\[-1mm]
&R_{\rm eq}=10.29^{+1.01}_{-0.86}\ \mathrm{km}.
\end{aligned}
\label{eq:nicer_points}
\end{equation}

These measurements test both the radius near $1.4M_\odot$ and the ability
of the model branch to remain stable beyond $2M_\odot$. They therefore
constrain the total EOS stiffness across different density intervals, not
one isolated pressure term.

PSR~J1231-1411 is another potentially useful target. However, two
independent interpretations of the same NICER/XMM-Newton data yield
substantially different radii: $M=1.04^{+0.05}_{-0.03}M_\odot$,
$R_{\rm eq}=12.6\pm0.3$~km in Ref.~\cite{SalmiJ1231_2024}, and
$M=1.12\pm0.07M_\odot$, $R_{\rm eq}=9.91^{+0.88}_{-0.86}$~km in the
independent analysis of Ref.~\cite{QiJ1231_2025}. This discrepancy reflects
sensitivity to hot-spot geometry, priors, and convergence of the pulse-
profile analysis. We therefore do not show J1231-1411 below as a single
cross-shaped marker; the two results are better used as an estimate of
model systematics in a future statistical test.

The comparison in Fig.~\ref{fig:CSC_MR} highlights a strength of the
baseline macroscopic model: with a compact parameter set it reproduces the
observed mass and radius scales and includes branches capable of supporting
masses near $2M_\odot$. Its original purpose was not a simultaneous Bayesian
fit to the current NICER set, so the added markers should be regarded as a
natural extension of the test, not as a rejection of the results of
Ref.~\cite{Uleiev2026}. PSR~J0437-4715 and especially the recent
PSR~J0614-3329 supply compact-radius benchmarks near $1.4M_\odot$, while
PSR~J0740+6620 tests whether the stable branch persists above $2M_\odot$.
Overlaying one-dimensional intervals is a useful diagnostic comparison; a
full statistical test with two-dimensional posteriors remains a separate
next step.

The proposed extension consistently includes explicit Fermi-liquid and
pairing corrections:
\begin{equation}
\begin{aligned}
P_{\mathrm{mod}}(\varepsilon)
={}&P_{\mathrm{ES}}(\varepsilon;K,C,a/R,\bar\rho)\\
&+\Delta P_{\mathrm{FL}}(\varepsilon;F_0^s,m^*)\\
&+\delta P_{\mathrm{pair}}(\varepsilon;\Delta),
\end{aligned}
\label{eq:mr_extended_eos}
\end{equation}
where $P_{\mathrm{ES}}$ denotes the baseline effective-surface description
of Ref.~\cite{Uleiev2026}. The term $\Delta P_{\mathrm{FL}}$ must be
understood as a \emph{residual} Landau correction after matching to the
mean-field interactions already encoded in $K$ and $C$. Simply adding the
full $P_{\mathrm{int}}^{(L)}$ to the baseline functional without such
matching would double count the interaction.

The parameter $F_0^s$ changes the compressibility and therefore the slope
and location of the $M(R)$ branch. Pairing adds a correction of a distinct
physical origin: for nucleonic matter its effect on global $M$ and $R$ is
usually small, whereas color superconductivity in a quark core can shift
the radius and maximum mass appreciably if the gap $\Delta$ is sufficiently
large. Joint calibration of $\Delta P_{\mathrm{FL}}$ and
$\delta P_{\mathrm{pair}}$ thus provides a physical mechanism that
\emph{may} improve simultaneous agreement with the four observational
points.

This is a testable hypothesis, not a guaranteed result: the sign and size
of the displacement depend on the density dependence of $F_0^s$, $m^*$,
and $\Delta$. Testing it requires reconstructing a causal and
thermodynamically stable $P(\varepsilon)$, solving the TOV equations for
each parameter set, and comparing the resulting curves with the full
two-dimensional NICER posteriors.

\FloatBarrier

\subsection{TOV equations for the Fermi-liquid region of a neutron star}
\label{sec:tov_fl}

Consider a cold, spherically symmetric neutron star whose interior is in
the degenerate Fermi-liquid regime,
\[
T\ll\varepsilon_F,\qquad n_B\lambda_T^3/g^*\gg1.
\]
The virial description is no longer applicable here, and the Pauli
principle manifests itself through the degeneracy pressure. Quasiparticle
interactions are incorporated through the effective mass, Landau
parameters, and the density dependence of the interaction energy.

We write the total pressure as
\begin{equation}
P(n_B)=P_F(n_B;m^*)+P_{\rm int}(n_B)
      +\delta P_{\rm pair}(n_B),
\label{eq:eos_total_fl}
\end{equation}
where $P_F$ is the degenerate quasiparticle pressure, $P_{\rm int}$ is the
strong-interaction contribution, and $\delta P_{\rm pair}$ is present when
$T<T_c$.

For a nonrelativistic nucleonic Fermi liquid,
\begin{equation}
k_F=(3\pi^2n_B)^{1/3},
\end{equation}
\begin{equation}
P_F(n_B;m^*)=
\frac{\hbar^2}{5m^*}(3\pi^2)^{2/3}n_B^{5/3}.
\label{eq:PF_nonrel}
\end{equation}

Within the Fermi-liquid parameterization, the quasiparticle interaction
may be represented by the same $F_0^s$ parameter and
$P_{\rm int}^{(L)}$ contribution defined in
Eqs.~\eqref{eq:landau_compressibility}--\eqref{eq:landau_pressure}.
If $m^*$ and $F_0^s$ vary slowly over the relevant density interval,
\begin{equation}
P_{\rm FL}\simeq(1+F_0^s)P_F,
\qquad
P_{\rm int}^{(L)}\simeq F_0^sP_F.
\label{eq:Pint_landau_constant}
\end{equation}
This approximation directly relates quasiparticle interactions to EOS
stiffness. To avoid double counting, $P_{\rm int}^{(L)}$ should be used as
an alternative to the phenomenological $E_{\rm int}$ parameterization
below, rather than adding both terms simultaneously.

The interaction term may instead be specified by the energy per baryon
$E_{\rm int}(n_B)$:
\begin{equation}
\begin{aligned}
\varepsilon(n_B)={}&n_Bm_Nc^2+\frac{3}{5}n_B\varepsilon_F^*(n_B;m^*)\\
&+n_BE_{\rm int}(n_B)+\delta\varepsilon_{\rm pair}(n_B),
\end{aligned}
\label{eq:energy_density_fl}
\end{equation}
where $\varepsilon_F^*=\hbar^2k_F^2/(2m^*)$ is the nonrelativistic
quasiparticle Fermi energy. The factor $3/5$ is the zero-temperature
mean kinetic energy per particle.
\begin{equation}
P_{\rm int}(n_B)=n_B^2\frac{dE_{\rm int}(n_B)}{dn_B}.
\label{eq:Pint}
\end{equation}
For example, one may use the phenomenological form
\begin{equation}
E_{\rm int}(n_B)
=A\left(\frac{n_B}{n_0}\right)
+B\left(\frac{n_B}{n_0}\right)^\sigma,
\label{eq:Eint_param}
\end{equation}
which gives
\begin{equation}
P_{\rm int}(n_B)=n_0\left[
A\left(\frac{n_B}{n_0}\right)^2
+\sigma B\left(\frac{n_B}{n_0}\right)^{\sigma+1}
\right].
\label{eq:Pint_param}
\end{equation}

In the BCS approximation, the pairing correction is
\begin{equation}
\delta P_{\rm pair}(n_B)
\simeq\frac{1}{2}N_{\rm pair}^*(0)\Delta^2(n_B)
=\frac{1}{4}N_{\rm tot}^*(0)\Delta^2(n_B),
\label{eq:Ppair_fl}
\end{equation}
where
\begin{equation}
N_{\rm pair}^*(0)=\frac{m^*k_F}{2\pi^2\hbar^2},
\qquad
N_{\rm tot}^*(0)=\frac{m^*k_F}{\pi^2\hbar^2}
\end{equation}
are respectively the single-spin (paired-channel) and total two-spin
densities of states at the Fermi surface. Writing the convention explicitly
removes the otherwise common factor-of-two ambiguity in the BCS
condensation energy.

The resulting EOS $P=P(\varepsilon)$ is inserted into the
Tolman--Oppenheimer--Volkoff equations,
\begin{equation}
\frac{dP}{dr}
=-
\frac{G}{c^2r^2}
\frac{\left[\varepsilon(r)+P(r)\right]
\left[m(r)+4\pi r^3P(r)/c^2\right]}
{1-\dfrac{2Gm(r)}{rc^2}},
\label{eq:TOV_P}
\end{equation}
\begin{equation}
\frac{dm}{dr}=4\pi r^2\frac{\varepsilon(r)}{c^2}.
\label{eq:TOV_m}
\end{equation}
The central boundary conditions are
\begin{equation}
m(0)=0,\qquad n_B(0)=n_c,\qquad P(0)=P(n_c).
\end{equation}
Integration proceeds outward until
\begin{equation}
P(R)=0,
\end{equation}
which defines the stellar radius $R$; the total mass is
\begin{equation}
M=m(R).
\end{equation}

Thus, the Fermi-liquid region affects the TOV solution through the
effective EOS stiffness,
\begin{equation}
\frac{dP}{d\varepsilon}=\frac{dP/dn_B}{d\varepsilon/dn_B}.
\label{eq:stiffness}
\end{equation}
This quantity controls the resistance to gravitational compression.
Increasing $P_{\rm int}$ or decreasing the effective mass $m^*$ stiffens
the EOS and therefore generally increases the stellar radius and maximum
mass.

\subsection{Pauli criterion near the relativistic degeneracy boundary}

The boundary between the classical and quantum-degenerate regimes may be
specified by the standard wave-packet overlap criterion,
\begin{equation}
\chi\equiv\frac{n_B\lambda_T^3}{g^*}\simeq1,
\label{eq:pauli_overlap}
\end{equation}
where $n_B$ is the number density of the effective component, $g^*$ is its
internal degeneracy, and $\lambda_T$ is the characteristic thermal de
Broglie wavelength. For a nonrelativistic gas,
\begin{equation}
\lambda_T=\sqrt{\frac{2\pi\hbar^2}{mT}}.
\end{equation}
This criterion fixes the onset only up to a factor of order unity; the
controlled virial domain is $\chi\ll1$. Near a relativistic degeneracy
boundary it is more precise to use the Fermi momentum obtained from the
density,
\begin{equation}
p_F=\hbar\left(\frac{6\pi^2n_B}{g^*}\right)^{1/3}.
\label{eq:pF_density}
\end{equation}
Including interactions, introduce the effective chemical potential
\begin{equation}
\mu^*(n_B,T)=\mu-U(n_B,T),
\label{eq:mu_star_interaction}
\end{equation}
where $U(n_B,T)$ is the interaction mean field.

The relativistic relation between the effective chemical potential and the
Fermi momentum is
\begin{equation}
\mu^*=\sqrt{p_F^2c^2+m^{*2}c^4},
\label{eq:mu_star_pF}
\end{equation}
so that
\begin{equation}
p_F=\frac{1}{c}\sqrt{\mu^{*2}-m^{*2}c^4}.
\label{eq:pF_mu_star}
\end{equation}

The Fermi temperature is defined by the kinetic energy of a quasiparticle
at the Fermi surface,
\begin{equation}
T_F^*=\sqrt{p_F^2c^2+m^{*2}c^4}-m^*c^2
=\mu^*-m^*c^2.
\label{eq:TF_mu_star}
\end{equation}
The onset of quantum degeneracy can therefore be written as
\begin{equation}
T\simeq T_F^*=\mu-U(n_B,T)-m^*c^2.
\label{eq:T_mu_boundary}
\end{equation}
Equations~\eqref{eq:pF_density} and \eqref{eq:TF_mu_star} give the explicitly
density-based relativistic boundary. In the ultrarelativistic limit,
\begin{equation}
T_F^*\simeq p_Fc
=\hbar c\left(\frac{6\pi^2n_B}{g^*}\right)^{1/3}
\simeq\mu^*,
\label{eq:ultrarel_pauli}
\end{equation}
where the last equality is valid only when $\mu^*\gg m^*c^2$.

For $T>T_F^*$ the system is classical or weakly degenerate, while for
$T<T_F^*$ it is a quantum-degenerate Fermi liquid. In the latter regime,
the Pauli principle is not a small virial correction but the primary source
of pressure:
\begin{equation}
P_{\rm Pauli}=\frac{g^*}{3}\int_0^{p_F}
\frac{d^3p}{(2\pi\hbar)^3}
\frac{p^2c^2}{\sqrt{p^2c^2+m^{*2}c^4}}.
\label{eq:P_pauli_rel_final}
\end{equation}
The full interacting Fermi-liquid pressure is
\begin{equation}
P(n_B)=P_{\rm Pauli}(n_B;m^*)+P_{\rm int}(n_B)
+\delta P_{\rm pair}(n_B).
\label{eq:P_total_pauli_interaction}
\end{equation}
Thus, the relativistic degeneracy boundary in an interacting system is not
simply $T\simeq\mu$, but
\[
T\simeq T_F^*=\mu-U(n_B,T)-m^*c^2,
\]
the kinetic Fermi energy of the quasiparticles. This scale determines the
location of the Fermi surface, the Pauli pressure, and the stiffness of the
EOS supplied to the TOV equations.

\subsection{Virial Pauli correction for quasiparticles near the Boltzmann boundary}

Near the boundary between the classical and quantum-degenerate regimes,
\begin{equation}
\chi^*\equiv\frac{n_B\lambda_T^{*3}}{g^*}\lesssim1,
\end{equation}
matter can still be described by a virial expansion, but for Fermi-liquid
quasiparticles rather than bare particles. Define
\begin{equation}
\mu^*=\mu-U(n_B,T)
\end{equation}
and the effective thermal de Broglie wavelength
\begin{equation}
\lambda_T^*=\sqrt{\frac{2\pi\hbar^2}{m^*T}}.
\label{eq:lambda_star_nonrel}
\end{equation}
The virial pressure is then
\begin{equation}
\frac{P}{T}=n_B+B_2^{\rm eff}(T,n_B)n_B^2+\mathcal O(n_B^3).
\label{eq:virial_pressure_star}
\end{equation}
We write the effective second virial coefficient as
\begin{equation}
B_2^{\rm eff}=B_2^{\rm int}+B_2^{\rm Pauli,*}.
\label{eq:B2_eff}
\end{equation}
For a van der Waals interaction,
\begin{equation}
B_2^{\rm int}(T)=b^*-\frac{a^*}{T},
\label{eq:B2_int}
\end{equation}
where $b^*$ represents effective short-range quasiparticle repulsion and
$a^*$ represents intermediate-range attraction. The Pauli correction for
fermionic quasiparticles is positive:
\begin{equation}
B_2^{\rm Pauli,*}(T)=\frac{\lambda_T^{*3}}{2^{5/2}g^*}.
\label{eq:B2_pauli_star_tov}
\end{equation}
Consequently,
\begin{equation}
B_2^{\rm eff}(T,n_B)=b^*-\frac{a^*}{T}
+\frac{1}{2^{5/2}g^*}
\left(\frac{2\pi\hbar^2}{m^*T}\right)^{3/2}.
\label{eq:B2_eff_final}
\end{equation}
Near the Boltzmann boundary, the Pauli principle thus acts as an additional
effective repulsion between quasiparticles. The pressure in this region is
\begin{equation}
P(n_B,T)=n_BT\left[1+B_2^{\rm eff}(T,n_B)n_B\right].
\label{eq:P_virial_pauli}
\end{equation}
The corresponding energy density may be written as
\begin{equation}
\begin{aligned}
\varepsilon(n_B,T)={}&n_Bm_Nc^2+\frac{3}{2}n_BT\\
&+\varepsilon_{\rm int}(n_B,T)+\varepsilon_{\rm Pauli}(n_B,T).
\end{aligned}
\label{eq:eps_virial_pauli}
\end{equation}
Thermodynamic consistency requires
\begin{equation}
\varepsilon_{\rm Pauli}=-T^2n_B^2
\frac{\partial B_2^{\rm Pauli,*}}{\partial T},
\end{equation}
including any temperature dependence of $m^*(T,n_B)$. If $m^*$ depends
only weakly on temperature,
\begin{equation}
\varepsilon_{\rm Pauli}\simeq
+\frac{3}{2}Tn_B^2B_2^{\rm Pauli,*}.
\label{eq:eps_pauli_simple}
\end{equation}
The EOS in the transition region may therefore be represented as
\begin{equation}
P=P(\varepsilon;T,m^*,a^*,b^*,g^*).
\label{eq:eos_transition}
\end{equation}
Equation~\eqref{eq:eos_transition} is inserted into the TOV system
\eqref{eq:TOV_P}--\eqref{eq:TOV_m}; the system itself need not be repeated.

In a numerical calculation, one evaluates at every stellar radius
\begin{equation}
B_2^{\rm eff}(r)=b^*(r)-\frac{a^*(r)}{T(r)}
+\frac{\lambda_T^{*3}(r)}{2^{5/2}g^*},
\end{equation}
and then uses
\begin{equation}
P(r)=n_B(r)T(r)\left[1+B_2^{\rm eff}(r)n_B(r)\right].
\label{eq:P_r_final}
\end{equation}
The positive term
\begin{equation}
\Delta B_2^{\rm Pauli,*}=\frac{\lambda_T^{*3}}{2^{5/2}g^*}
\end{equation}
increases the pressure at fixed density and therefore stiffens the EOS. In
the TOV equations, this increases the resistance to gravitational
compression and can raise the stellar radius and maximum mass if the
transition region occupies a non-negligible fraction of the star.

\begin{figure*}[!t]
\centering
\includegraphics[width=0.96\textwidth]{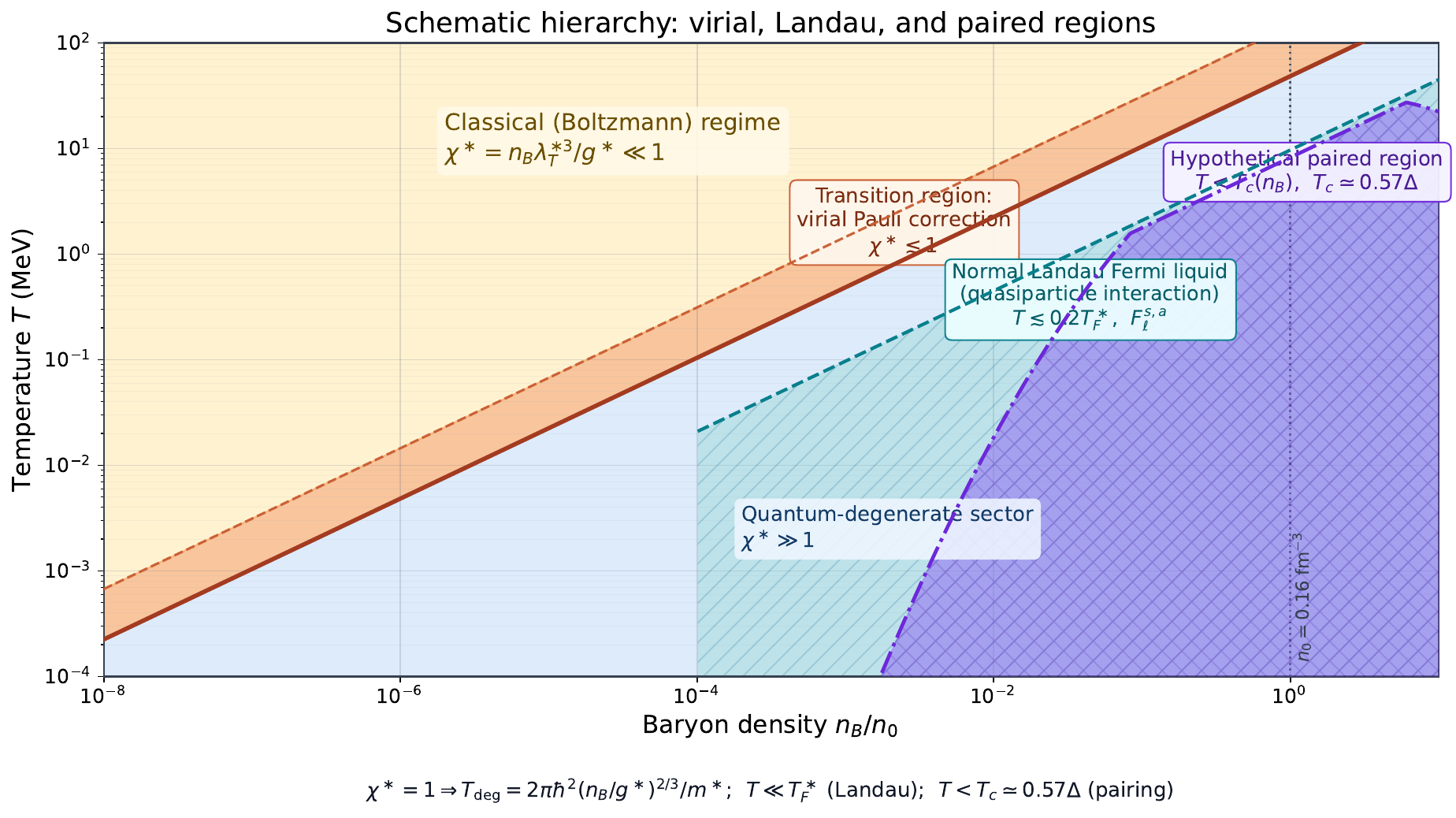}
\caption{Schematic hierarchy of regimes in the $(T,n_B)$ plane. The solid
brown line is the criterion $\chi^*=n_B\lambda_T^{*3}/g^*=1$ and, for a
transparent one-component normalization, is drawn with $m^*=m_N$ and
$g^*=2$; its scaling $T_{\rm deg}\propto(n_B/g^*)^{2/3}/m^*$ follows directly from
Eq.~\eqref{eq:lambda_star_nonrel}. The orange band schematically denotes
the transition regime $\chi^*\lesssim1$, in which the Pauli
term is added to the second virial coefficient. The classical Boltzmann
regime lies above the band and quantum-degenerate matter below it. Turquoise
hatching approximately marks the normal Landau-Fermi-liquid region, where
$T\ll T_F^*$ and quasiparticle interactions are described by the parameters
$F_\ell^{s,a}$; the illustrative boundary $T=0.2T_F^*$ is not a phase
transition. The violet region is a \emph{hypothetical} pairing domain
$T<T_c(n_B)$, with its upper boundary schematically tied to the weak-
coupling relation $T_c\simeq0.57\Delta$. The shape of this ``dome'' is not a
fit or a microscopic calculation; it only indicates that the paired phase
must be nested within the low-temperature degenerate region. Its actual
location depends on $m^*(n_B,T)$, the Landau parameters, the density
dependence of $\Delta(n_B)$, and the selected EOS.}
\label{fig:phase_transition_Tn}
\end{figure*}

The transition region is most reliably displayed directly in the
$(T,n_B)$ plane, because the criterion
$\chi^*=n_B\lambda_T^{*3}/g^*\sim1$ is formulated in these variables.
A representation in the $(T,\mu_B)$ plane
would require an additional model-dependent relation $n_B(T,\mu_B)$, that
is, a specific EOS. Figure~\ref{fig:phase_transition_Tn} therefore uses the
density representation only. Within the quantum-degenerate sector it also
shows the approximate normal Landau-Fermi-liquid region and a hypothetical
pairing region nested inside it.

As shown schematically in Fig.~\ref{fig:phase_transition_Tn}, the modified
second virial coefficient in Eq.~\eqref{eq:B2_eff_final} applies only in a
narrow transition region near the Boltzmann boundary
$n_B\lambda_T^{*3}/g^*\lesssim1$, where Fermi-liquid quasiparticles already feel
Pauli repulsion but the system has not yet become fully quantum degenerate.
The correction is small in the classical region, whereas deep in the
degenerate region the virial expansion must be replaced by a Fermi-liquid
EOS. In the turquoise region, the pressure contains the explicit Landau
term $P_{\rm int}^{(L)}$; below the hypothetical $T_c(n_B)$ line, the normal
Fermi-liquid description is supplemented by $\delta P_{\rm pair}$.

\begin{figure*}[!t]
\centering
\includegraphics[width=0.90\textwidth]{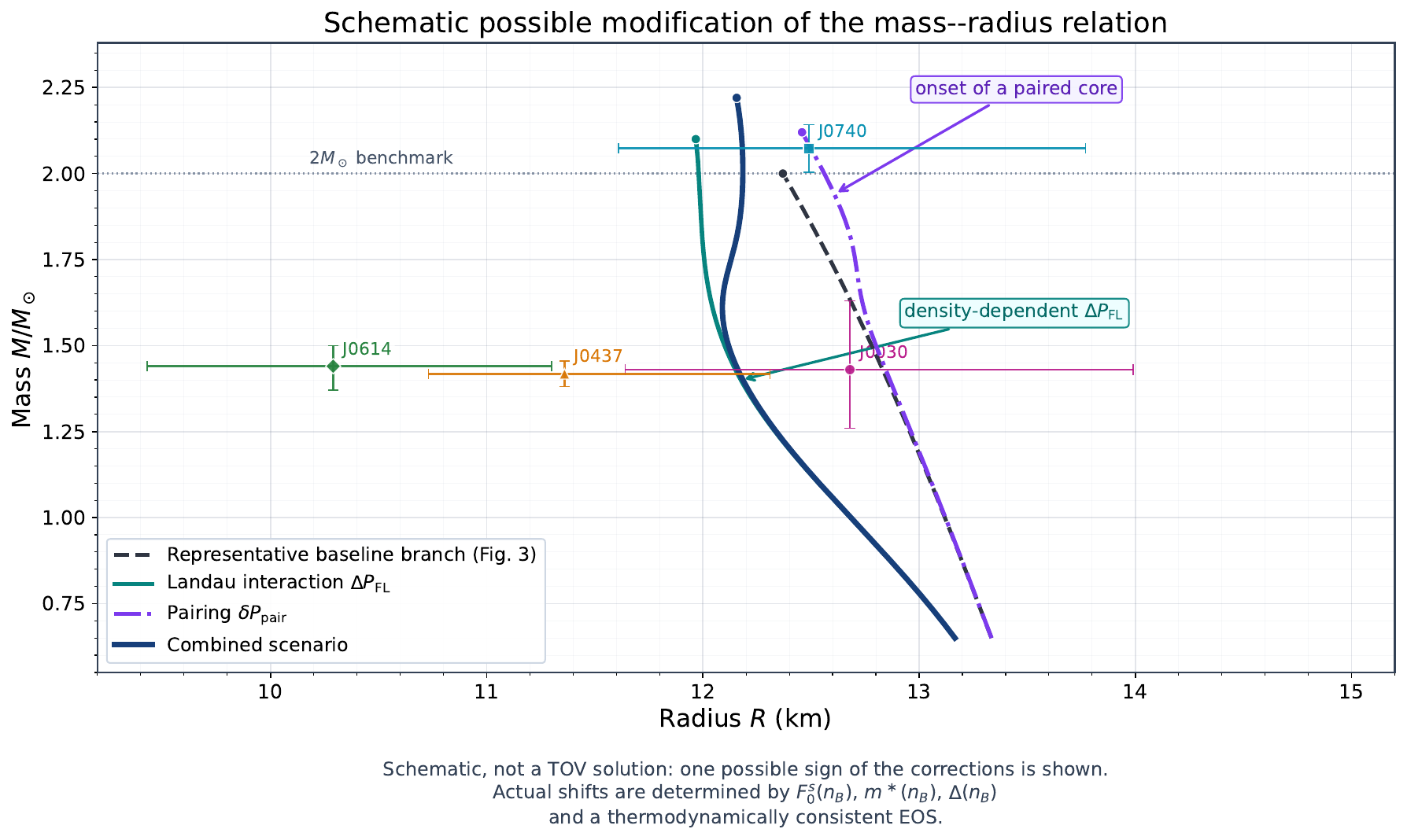}
\caption{Schematic illustration of a possible modification of the
representative $M(R)$ branch in Fig.~\ref{fig:CSC_MR}. The black dashed
curve denotes a nominal baseline branch; it reproduces only the typical
scale of the source figure and is not a digitization. The turquoise curve
illustrates a change of the entire branch caused by a density-dependent
residual Landau correction $\Delta P_{\rm FL}$. The violet dash-dotted
curve shows a scenario in which $\delta P_{\rm pair}$ becomes visible after
a paired core appears and mainly changes the high-mass part of the sequence.
The dark-blue curve shows their possible combined action. For illustration,
we choose a scenario in which $\Delta P_{\rm FL}$ slightly reduces the
radius near $1.4M_\odot$, while the positive pairing term supports a larger
maximum mass. This is neither a numerical TOV solution nor a statistical
fit: the sign and magnitude of the shifts depend on $F_0^s(n_B)$,
$m^*(n_B)$, $\Delta(n_B)$, and a thermodynamically consistent EOS. Colored
markers reproduce the same one-dimensional projections of the $68\%$
NICER intervals as in Fig.~\ref{fig:CSC_MR}, based on the primary analyses
\cite{Kini2026,Salmi2024,Choudhury2024,Mauviard2025}.}
\label{fig:MR_landau_pairing_schematic}
\end{figure*}

A possible qualitative consequence of these two corrections for the
mass--radius relation is shown in
Fig.~\ref{fig:MR_landau_pairing_schematic}. Since the Landau interaction
operates throughout the degenerate sector, it can displace a substantial
part of the $M(R)$ branch. Pairing instead affects mainly the part of the
sequence in which a paired inner core has already formed.

Figure~\ref{fig:MR_landau_pairing_schematic} thus visualizes the hypothesis
of Eq.~\eqref{eq:mr_extended_eos}: the Landau correction can control the
characteristic radius, while pairing modifies the high-density termination
of the branch and its maximum mass. The apparent approach of the combined
curve to some NICER central values should not be interpreted as a fit. The
new compact-radius benchmark J0614-3329 appreciably broadens the range over
which the magnitude and density dependence of the corrections should be
calibrated in a joint quantitative analysis. Such a test requires an
explicit $P_{\rm mod}(\varepsilon)$ followed by a new TOV integration.

In the scenario shown, the dominant displacement of the $M(R)$ relation is
produced by the residual Landau quasiparticle interaction
$\Delta P_{\rm FL}$, whereas the pairing correction
$\delta P_{\rm pair}$ is a subleading high-density contribution that
primarily modifies the termination of the stable branch and the maximum
stellar mass. This hierarchy is consistent with estimates in which pairing
corrects the degenerate quark pressure, while quasiparticle interactions
enter the bulk EOS throughout the Fermi-liquid region
\cite{Alp2025,ShabalBachelor2026}.
This dominance is an assumption built into the schematic scenario of
Fig.~5, not a result derived from a numerical TOV solution.

\subsection{TOV solutions with the Pauli correction to the second virial coefficient}

Consider a spherically symmetric compact star whose equilibrium is governed
by the TOV system already given in
Eqs.~\eqref{eq:TOV_P}--\eqref{eq:TOV_m}. We now focus only on the additional
Pauli correction to the EOS in the transition region.

For $n_B\lambda_T^{*3}/g^*\lesssim1$, we use the virial EOS
\eqref{eq:P_virial_pauli} with the coefficient
\eqref{eq:B2_eff_final}. For an analytic estimate, it is useful to separate
the Pauli part of the pressure:
\begin{equation}
P(n_B,T)=P_0(n_B,T)+\delta P_{\rm Pauli}(n_B,T),
\label{eq:P_split}
\end{equation}
where
\begin{equation}
P_0(n_B,T)=n_BT\left[1+\left(b^*-\frac{a^*}{T}\right)n_B\right],
\end{equation}
and
\begin{equation}
\delta P_{\rm Pauli}=n_B^2T\frac{\lambda_T^{*3}}{2^{5/2}g^*}.
\label{eq:deltaP_pauli}
\end{equation}

In the simplest approximation, assume that the temperature and effective
mass vary slowly across the narrow transition region. Then
\begin{equation}
\delta P_{\rm Pauli}\simeq K_{\rm P}n_B^2,
\label{eq:deltaP_KP}
\end{equation}
with
\begin{equation}
K_{\rm P}=T\frac{\lambda_T^{*3}}{2^{5/2}g^*}
=\frac{T}{2^{5/2}g^*}
\left(\frac{2\pi\hbar^2}{m^*T}\right)^{3/2}.
\label{eq:KP_def}
\end{equation}
The pressure can be represented as an effective mixed polytrope,
but the controlled virial expansion itself does not permit one to assume
that its quadratic term is parametrically larger than the ideal term.
For a purely analytic sensitivity estimate, we instead approximate the
background EOS locally by a $\Gamma=2$ polytrope,
\begin{equation}
P_0(n_B)\simeq K_0n_B^2,
\label{eq:P_poly_mixed}
\end{equation}
and add the Pauli term in Eq.~\eqref{eq:deltaP_KP}. This gives
\begin{equation}
P=K_{\rm eff}n_B^2,
\qquad
K_{\rm eff}=K_0
+\frac{T\lambda_T^{*3}}{2^{5/2}g^*}.
\label{eq:Keff}
\end{equation}
Here $K_0$ summarizes the local degenerate/interacting background and is
not identified with the second-order virial coefficient. This construction
is a formal response estimate, not an extrapolation of the virial EOS to
the entire star.
In terms of the mass density $\rho=m_Bn_B$, this becomes
\begin{equation}
P=\widetilde K_{\rm eff}\rho^2,
\qquad
\widetilde K_{\rm eff}=\frac{K_{\rm eff}}{m_B^2}.
\label{eq:polytrope_rho}
\end{equation}

For a polytrope with $\Gamma=2$, the Newtonian limit of the TOV equations
reduces to the Lane--Emden equation with index $N=1$. Its solution is
\begin{equation}
\theta(\xi)=\frac{\sin\xi}{\xi},
\label{eq:lane_emden_solution}
\end{equation}
where
\begin{equation}
\rho(r)=\rho_c\theta(\xi),
\qquad r=\alpha\xi,
\end{equation}
and
\begin{equation}
\alpha^2=\frac{\widetilde K_{\rm eff}}{2\pi G}.
\label{eq:alpha_def}
\end{equation}
The stellar radius is set by the first zero of $\theta(\xi)$,
\begin{equation}
\xi_1=\pi,
\qquad R=\pi\alpha,
\end{equation}
so that
\begin{equation}
R=\pi\left(\frac{\widetilde K_{\rm eff}}{2\pi G}\right)^{1/2}.
\label{eq:R_pauli}
\end{equation}
The configuration mass is
\begin{equation}
M=4\pi^2\alpha^3\rho_c.
\label{eq:M_pauli}
\end{equation}

The Pauli correction to the second virial coefficient therefore changes
the effective EOS stiffness,
\begin{equation}
\widetilde K_{\rm eff}=\widetilde K_0
+\delta\widetilde K_{\rm Pauli},
\end{equation}
where
\begin{equation}
\delta\widetilde K_{\rm Pauli}=\frac{1}{m_B^2}
\frac{T\lambda_T^{*3}}{2^{5/2}g^*}.
\label{eq:deltaK_pauli}
\end{equation}
In this approximation, the relative radius change is
\begin{equation}
\frac{\delta R}{R}\simeq\frac{1}{2}
\frac{\delta\widetilde K_{\rm Pauli}}{\widetilde K_0},
\label{eq:deltaR}
\end{equation}
whereas at fixed central density $\rho_c$ the relative mass change is
\begin{equation}
\frac{\delta M}{M}\simeq\frac{3}{2}
\frac{\delta\widetilde K_{\rm Pauli}}{\widetilde K_0}.
\label{eq:deltaM}
\end{equation}

Thus, in the simplest approximation the Pauli contribution to the second
virial coefficient produces a positive correction to the effective
polytropic constant,
\begin{equation}
K_{\rm eff}=K_0+\frac{T\lambda_T^{*3}}{2^{5/2}g^*}.
\end{equation}
Because this term is positive, it stiffens the EOS. In TOV solutions this
appears as an increase of the characteristic radius and, at fixed central
density, of the configuration mass. Equations~\eqref{eq:deltaR} and
\eqref{eq:deltaM} quantify only the sign and local sensitivity within the
auxiliary $N=1$ polytropic model; a physical prediction requires matching
the virial layer to a complete causal EOS and integrating the TOV equations
numerically.

\FloatBarrier

\section*{Declaration of generative AI and AI-assisted technologies}

During the preparation of this manuscript the authors used ChatGPT (OpenAI) solely as an assistive tool for English-language editing, structural suggestions, literature-search assistance, code debugging and testing, numerical cross-checks, and preparation of figures from the authors’ own data. All physical assumptions, model definitions, objective functions, data selection, interpretation of results, and final conclusions were made by the human authors. Every cited source and all numerical results were independently verified by the authors. The authors take full responsibility for the content of the published article.

\section{Conclusions}

We have developed a consistent extension of the van der Waals equation of
state \cite{Krivenko2017Attraction,Krivenko2019Attraction,Vovchenko2017,Krivenko2023}
in which pressure contributions are ordered according to the hierarchy
\begin{equation}
P_{\mathrm{Pauli}}
\rightarrow
P_{\mathrm{interaction}}
\rightarrow
P_{\mathrm{pairing}}.
\end{equation}

In the dilute regime, the Pauli principle gives a positive temperature-
dependent correction to the second virial coefficient; in degenerate
matter, it appears through the dominant Fermi pressure. In the normal
phase, quasiparticle interactions are parameterized by $F_0^s$, explicitly
linking microscopic interactions with the compressibility and sound speed.

Pairing correlations provide a positive pressure correction at fixed
chemical potential. Their effect on the mass--radius relation is usually
small in nucleonic matter, whereas color superconductivity in a quark core
can reach the $10$--$20\%$ level for large gaps
\cite{Alford2008,Kurkela2024}.

The Tolman--Oppenheimer--Volkoff equations translate this microscopic
hierarchy into the stellar mass and radius. In the simplest polytropic
estimate, the Pauli correction near the Boltzmann boundary increases the
effective stiffness and therefore the characteristic radius and mass.

Comparison with four NICER results
\cite{Kini2026,Salmi2024,Choudhury2024,Mauviard2025} shows that an
acceptable model branch must be tested against both the compact benchmark
$R\simeq10$--$11$~km near $1.4M_\odot$ from PSR~J0614-3329 and the larger
radii of other sources, while simultaneously supporting configurations of
at least $2M_\odot$ \cite{Antoniadis2013,Salmi2024}. In this respect, the macroscopic approach of
Ref.~\cite{Uleiev2026} provides a productive and transparent baseline for
further microscopic refinement.


@article{Fukushima2011,
  author   = {Fukushima, Kenji and Hatsuda, Tetsuo},
  title    = {The phase diagram of dense {QCD}},
  journal  = {Reports on Progress in Physics},
  year     = {2011},
  volume   = {74},
  number   = {1},
  pages    = {014001},
  doi      = {10.1088/0034-4885/74/1/014001},
  langid   = {english}
}

@article{Krivenko2023,
  author   = {Krivenko-Emetov, Yaroslav D.},
  title    = {Multicomponent Van der Waals Model of a Nuclear Fireball in the Freeze-Out Stage},
  journal  = {Letters in High Energy Physics},
  year     = {2023},
  volume   = {2023},
  pages    = {401},
  doi      = {10.31526/lhep.2023.401},
  eprint   = {2301.00742},
  eprinttype = {arxiv},
  eprintclass = {hep-ph},
  langid   = {english}
}

@online{Krivenko2019Attraction,
  author   = {Krivenko-Emetov, Yaroslav D.},
  title    = {Attractive Inter-Particle Force in Van der Waals Model of Hadron Gas in the Grand Canonical Ensemble},
  year     = {2019},
  eprint   = {1909.08441},
  eprinttype = {arxiv},
  eprintclass = {hep-ph},
  url      = {https://arxiv.org/abs/1909.08441},
  urldate  = {2026-08-09},
  langid   = {english}
}

@article{Vovchenko2017,
  author   = {Vovchenko, Volodymyr and Motornenko, Anton and Alba, Paolo and Gorenstein, Mark I. and Satarov, Leonid M. and Stoecker, Horst},
  title    = {Multicomponent van der Waals equation of state: Applications in nuclear and hadronic physics},
  journal  = {Physical Review C},
  year     = {2017},
  volume   = {96},
  pages    = {045202},
  doi      = {10.1103/PhysRevC.96.045202},
  eprint   = {1707.09215},
  eprinttype = {arxiv},
  langid   = {english}
}

@article{Tolman1939,
  author   = {Tolman, Richard C.},
  title    = {Static Solutions of Einstein's Field Equations for Spheres of Fluid},
  journal  = {Physical Review},
  year     = {1939},
  volume   = {55},
  pages    = {364--373},
  doi      = {10.1103/PhysRev.55.364},
  langid   = {english}
}

@article{Oppenheimer1939,
  author   = {Oppenheimer, J. Robert and Volkoff, George M.},
  title    = {On Massive Neutron Cores},
  journal  = {Physical Review},
  year     = {1939},
  volume   = {55},
  pages    = {374--381},
  doi      = {10.1103/PhysRev.55.374},
  langid   = {english}
}

@article{Lattimer2001,
  author   = {Lattimer, James M. and Prakash, Madappa},
  title    = {Neutron Star Structure and the Equation of State},
  journal  = {The Astrophysical Journal},
  year     = {2001},
  volume   = {550},
  pages    = {426--442},
  doi      = {10.1086/319702},
  langid   = {english}
}

@article{Antoniadis2013,
  author   = {Antoniadis, John and Freire, Paulo C. C. and Wex, Norbert and Tauris, Thomas M. and Lynch, Ryan S. and van Kerkwijk, Marten H. and others},
  title    = {A Massive Pulsar in a Compact Relativistic Binary},
  journal  = {Science},
  year     = {2013},
  volume   = {340},
  pages    = {613--616},
  doi      = {10.1126/science.1233232},
  eprint   = {1304.6875},
  eprinttype = {arxiv},
  eprintclass = {astro-ph.HE},
  langid   = {english}
}

@book{LandauLifshitz,
  author    = {Landau, L. D. and Lifshitz, E. M.},
  title     = {Statistical Physics, Part 1},
  series    = {Course of Theoretical Physics},
  volume    = {5},
  edition   = {3},
  publisher = {Butterworth-Heinemann},
  year      = {1980},
  langid    = {english}
}

@article{BCS1957,
  author   = {Bardeen, John and Cooper, L. N. and Schrieffer, J. R.},
  title    = {Theory of Superconductivity},
  journal  = {Physical Review},
  year     = {1957},
  volume   = {108},
  pages    = {1175--1204},
  doi      = {10.1103/PhysRev.108.1175},
  langid   = {english}
}

@article{Alford2008,
  author   = {Alford, Mark G. and Schmitt, Andreas and Rajagopal, Krishna and Schaefer, Thomas},
  title    = {Color superconductivity in dense quark matter},
  journal  = {Reviews of Modern Physics},
  year     = {2008},
  volume   = {80},
  pages    = {1455--1515},
  doi      = {10.1103/RevModPhys.80.1455},
  eprint   = {0709.4635},
  eprinttype = {arxiv},
  langid   = {english}
}

@article{Kurkela2024,
  author   = {Kurkela, Aleksi and Rajagopal, Krishna and Steinhorst, Rachel},
  title    = {Astrophysical Equation-of-State Constraints on the Color-Superconducting Gap},
  journal  = {Physical Review Letters},
  year     = {2024},
  volume   = {132},
  pages    = {262701},
  doi      = {10.1103/PhysRevLett.132.262701},
  eprint   = {2401.16253},
  eprinttype = {arxiv},
  eprintclass = {hep-ph},
  langid   = {english}
}

@inproceedings{Krivenko2017Attraction,
  author    = {Krivenko-Emetov, Yaroslav D.},
  title     = {Interparticle attractive forces account of the multicomponent hadron gas in the grand canonical ensemble},
  booktitle = {Book of Abstracts of the XXIV Annual Scientific Conference of the Institute for Nuclear Research of the National Academy of Sciences of Ukraine},
  year      = {2017},
  location  = {Kyiv},
  pages     = {36},
  publisher = {Institute for Nuclear Research, NAS of Ukraine},
  url       = {https://inis.iaea.org/records/cnyyh-3vr91},
  urldate   = {2026-08-09},
  note      = {In Ukrainian},
  langid    = {ukrainian}
}

@article{Alp2025,
  author   = {Alp, Faruk and Dietz, Yannick and Hebeler, Kai and Schwenk, Achim},
  title    = {Equation of state and Fermi liquid properties of dense matter based on chiral effective field theory interactions},
  journal  = {Physical Review C},
  year     = {2025},
  volume   = {112},
  pages    = {055802},
  doi      = {10.1103/ls3l-dn1y},
  eprint   = {2504.18259},
  eprinttype = {arxiv},
  langid   = {english}
}

@article{Schwenk2003,
  author   = {Schwenk, Achim and Friman, Bengt and Brown, Gerald E.},
  title    = {Renormalization Group Approach to Neutron Matter: Quasiparticle Interactions, Superfluid Gaps and the Equation of State},
  journal  = {Nuclear Physics A},
  year     = {2003},
  volume   = {713},
  pages    = {191--216},
  doi      = {10.1016/S0375-9474(02)01290-3},
  eprint   = {nucl-th/0207004},
  eprinttype = {arxiv},
  langid   = {english}
}

@article{SedrakianClark2019,
  author   = {Sedrakian, Armen and Clark, John W.},
  title    = {Superfluidity in Nuclear Systems and Neutron Stars},
  journal  = {European Physical Journal A},
  year     = {2019},
  volume   = {55},
  pages    = {167},
  doi      = {10.1140/epja/i2019-12863-6},
  eprint   = {1802.00017},
  eprinttype = {arxiv},
  eprintclass = {nucl-th},
  langid   = {english}
}

@article{Shternin2021,
  author   = {Shternin, Peter S. and Ofengeim, Dmitry D. and Ho, Wynn C. G. and Heinke, Craig O. and Wijngaarden, M. J. P. and Patnaude, Daniel J.},
  title    = {Model-Independent Constraints on Superfluidity from the Cooling Neutron Star in Cassiopeia A},
  journal  = {Monthly Notices of the Royal Astronomical Society},
  year     = {2021},
  volume   = {506},
  number   = {1},
  pages    = {709--726},
  doi      = {10.1093/mnras/stab1695},
  eprint   = {2106.05692},
  eprinttype = {arxiv},
  eprintclass = {astro-ph.HE},
  langid   = {english}
}

@online{Kini2026,
  author   = {Kini, Yves and Mauviard, Lucien and Salmi, Tuomo and Watts, Anna L. and Guillot, Sebastien and Dorsman, Bas and Choudhury, Devarshi and others},
  title    = {A NICER View of PSR J0030+0451: Updated Constraints from Six Years of NICER Observations},
  year     = {2026},
  eprint   = {2602.23743},
  eprinttype = {arxiv},
  eprintclass = {astro-ph.HE},
  url      = {https://arxiv.org/abs/2602.23743},
  urldate  = {2026-08-09},
  note     = {Submitted to The Astrophysical Journal},
  langid   = {english}
}

@article{Salmi2024,
  author   = {Salmi, Tuomo and Choudhury, Devarshi and Kini, Yves and Riley, Thomas E. and Vinciguerra, Serena and Watts, Anna L. and others},
  title    = {The Radius of the High-Mass Pulsar PSR J0740+6620 with 3.6 yr of NICER Data},
  journal  = {The Astrophysical Journal},
  year     = {2024},
  volume   = {974},
  number   = {2},
  pages    = {294},
  doi      = {10.3847/1538-4357/ad5f1f},
  eprint   = {2406.14466},
  eprinttype = {arxiv},
  eprintclass = {astro-ph.HE},
  langid   = {english}
}

@article{Choudhury2024,
  author   = {Choudhury, Devarshi and Salmi, Tuomo and Vinciguerra, Serena and Riley, Thomas E. and Kini, Yves and Watts, Anna L. and others},
  title    = {A NICER View of the Nearest and Brightest Millisecond Pulsar: PSR J0437-4715},
  journal  = {The Astrophysical Journal Letters},
  year     = {2024},
  volume   = {971},
  number   = {1},
  pages    = {L20},
  doi      = {10.3847/2041-8213/ad5a6f},
  eprint   = {2407.06789},
  eprinttype = {arxiv},
  eprintclass = {astro-ph.HE},
  langid   = {english}
}

@online{Dexheimer2026,
  author   = {Dexheimer, Veronica},
  title    = {The equation of state for neutron stars},
  year     = {2026},
  eprint   = {2607.17006},
  eprinttype = {arxiv},
  eprintclass = {nucl-th},
  url      = {https://arxiv.org/abs/2607.17006},
  urldate  = {2026-08-09},
  langid   = {english}
}

@article{Uleiev2026,
  author   = {Uleiev, A. A. and Magner, A. G. and Maydanyuk, S. P. and Bonasera, A. and Zheng, H. and Fedotkin, S. N. and Levon, A. I. and Grygoriev, U. V. and Depastas, T.},
  title    = {Macroscopic Approaches to Rotating Neutron Stars},
  journal  = {Nuclear Physics and Atomic Energy},
  year     = {2026},
  volume   = {27},
  number   = {1},
  pages    = {5--15},
  doi      = {10.15407/jnpae2026.01.005},
  langid   = {english}
}

@article{Mauviard2025,
  author   = {Mauviard, Lucien and Guillot, Sebastien and Salmi, Tuomo and Choudhury, Devarshi and Dorsman, Bas and González-Caniulef, Denis and others},
  title    = {A NICER View of the 1.4 Solar-Mass Edge-on Pulsar PSR J0614-3329},
  journal  = {The Astrophysical Journal},
  year     = {2025},
  volume   = {995},
  number   = {1},
  pages    = {60},
  doi      = {10.3847/1538-4357/ae145d},
  eprint   = {2506.14883},
  eprinttype = {arxiv},
  eprintclass = {astro-ph.HE},
  langid   = {english}
}

@thesis{ShabalBachelor2026,
  author      = {Shabal, Hlib Romanovych},
  title       = {Some Thermodynamic Properties of Hadronic Matter under Extreme Conditions},
  type        = {Bachelor's thesis},
  institution = {National Technical University of Ukraine ``Igor Sikorsky Kyiv Polytechnic Institute'', Institute of Physics and Technology},
  location    = {Kyiv},
  year        = {2026},
  note        = {Original in Ukrainian;  consultant: Y.~D.~Krivenko-Emetov},
  url         = {https://ela.kpi.ua/items/83632743-90f7-4a42-b6c8-979d1b7f5e57},
  urldate     = {2026-08-09},
  langid      = {english}
}

@article{SalmiJ1231_2024,
  author   = {Salmi, Tuomo and Deneva, Julia S. and Ray, Paul S. and Watts, Anna L. and Choudhury, Devarshi and Kini, Yves and others},
  title    = {A NICER View of PSR J1231-1411: A Complex Case},
  journal  = {The Astrophysical Journal},
  year     = {2024},
  volume   = {976},
  number   = {1},
  pages    = {58},
  doi      = {10.3847/1538-4357/ad81d2},
  eprint   = {2409.14923},
  eprinttype = {arxiv},
  eprintclass = {astro-ph.HE},
  langid   = {english}
}

@article{QiJ1231_2025,
  author   = {Qi, Liqiang and Zheng, Shijie and Zhang, Juan and Ge, Mingyu and Li, Ang and Zhang, Shuang-Nan and others},
  title    = {PSR J1231-1411 Revisited: Pulse Profile Analysis of X-Ray Observation},
  journal  = {The Astrophysical Journal},
  year     = {2025},
  volume   = {981},
  number   = {2},
  pages    = {99},
  doi      = {10.3847/1538-4357/adb42f},
  eprint   = {2502.09147},
  eprinttype = {arxiv},
  eprintclass = {astro-ph.HE},
  langid   = {english}
}
\end{document}